# An efficient steady-state analysis of the eddy current problem using a parallel-in-time algorithm

*I. Kulchytska-Ruchka\*, H. De Gersem\*, and S. Schöps\**

*\*Institut für Teilchenbeschleinigung und Elektromagnetische Felder (TEMF), Technische Universität Darmstadt, Schlossgartenstrasse 8, D-64289 Darmstadt, Germany; email: kulchytska@gsc.tu-darmstadt.de*

**Keywords:** Steady state, eddy currents, parallel-in-time algorithms, harmonic balance method, fast Fourier transform.

## Abstract

This paper introduces a parallel-in-time algorithm for efficient steady-state solution of the eddy current problem. Its main idea is based on the application of the well-known multi-harmonic (or harmonic balance) approach as the coarse solver within the periodic parallel-in-time framework. A frequency domain representation allows for the separate calculation of each harmonic component in parallel and therefore accelerates the solution of the time-periodic system. The presented approach is verified for a nonlinear coaxial cable model.

## 1 Introduction

Steady-state analysis of electro-mechanical appliances is often of big interest in engineering, especially during initial design stages. The classical time-stepping approach might require the sequential calculation of a large number of steps over many periods till the transient phenomenon subsides and the response becomes periodic. This is usually the case for systems whose time constants are much larger than the period of the driving force, as it is illustrated in Fig. 1 for an induction machine model [1, 2]. Various approaches for an efficient calculation of the steady state can be encountered in literature, e.g., the time-periodic explicit error correction method [3, 4].

The Parareal algorithm [5, 6] is a rather new and efficient parallel-in-time approach, which allows to speed up sequential time stepping via parallelization along the time axis. It was recently applied to the eddy current problem of an induction motor in [2], while [7] introduced a novel Parareal variant, able to treat systems excited with pulse-width-modulated input. The original Parareal approach was extended to the class of time-periodic problems in [8] using two algorithms: PP-IC (periodic Parareal algorithm with initial-value coarse problem) and PP-PC (periodic Parareal algorithm with periodic coarse problem). The first method imposes a relaxed periodicity, parallelizing the time stepping period by period until the steady state is achieved, and was applied to the induction motor in [9]. In contrast to this, PP-PC requires the solution of the time-periodic system over the whole period, discretized on a coarse grid. Solution of this system might be prohibitively expensive, since its size is usually quite large and the system matrix has a special block-cyclic structure.

The current paper uses the main idea of PP-PC, which we describe in Section 2 for the time-periodic eddy current problem, and reduces computational cost of the PP-PC system solution via introducing an additional parallelizable correction on the coarse grid. Section 3 presents our new multi-harmonic approach, applied to the time-periodic system of the PP-PC iteration. The performance of the introduced algorithm is shown in Section 4 through its application to a simple nonlinear coaxial cable model. We finally conclude our contribution in Section 5.

## 2 PP-PC for the eddy current problem

Consider the space-discrete time-periodic eddy current problem in A-formulation [10]

$$\mathbf{M}d_t\mathbf{u}(t) + \mathbf{K}(\mathbf{u}(t))\mathbf{u}(t) = \mathbf{f}(t), \quad t \in (0, T), \\ \mathbf{u}(0) = \mathbf{u}(T), \quad (1)$$

where $\mathbf{u}$ denotes the discretized magnetic vector potential, $\mathbf{M}$ the (singular) conductivity matrix, $\mathbf{K}$ is the curl-curl matrix, and $\mathbf{f}$ is defined by a given excitation. Parareal [5] is based on splitting of the interval $[0, T]$ into $N$ subintervals $[T_{n-1}, T_n]$, $n = 1, \ldots, N$, with $T_0 = 0$ and $T_N = T$, followed by coarse and fine grid solutions. Let

$$\mathcal{G}_n^k = \mathcal{G}_n\left(\mathbf{U}_{n-1}^k\right) \quad \text{and} \quad \mathcal{F}_n^k = \mathcal{F}_n\left(\mathbf{U}_{n-1}^k\right) \quad (2)$$

respectively denote the coarse and the fine solutions of the initial-value problem

$$\mathbf{M}d_t\mathbf{u}(t) + \mathbf{K}(\mathbf{u}(t))\mathbf{u}(t) = \mathbf{f}(t), \quad t \in (T_{n-1}, T_n], \\ \mathbf{u}(T_{n-1}) = \mathbf{U}_{n-1}^k, \quad (3)$$

at the end of the subinterval. The classical Parareal iteration [6] for $n = 1, \ldots, N$ together with the periodicity condition gives the PP-PC iteration [8], for $k = 0, 1, \ldots, K$

$$\mathbf{U}_0^{k+1} = \mathcal{F}_N\left(\mathbf{U}_{N-1}^k\right) + \mathcal{G}_N\left(\mathbf{U}_{N-1}^{k+1}\right) - \mathcal{G}_N\left(\mathbf{U}_{N-1}^k\right), \\ \mathbf{U}_n^{k+1} = \mathcal{F}_n\left(\mathbf{U}_{n-1}^k\right) + \mathcal{G}_n\left(\mathbf{U}_{n-1}^{k+1}\right) - \mathcal{G}_n\left(\mathbf{U}_{n-1}^k\right). \quad (4)$$

Note that the PP-IC method [8] is derived by relaxing the update of the initial value in (4)

$$\mathbf{U}_0^{k+1} = \mathbf{U}_N^k. \quad (5)$$

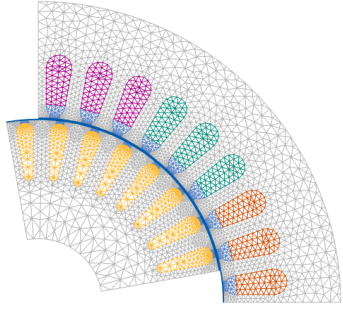 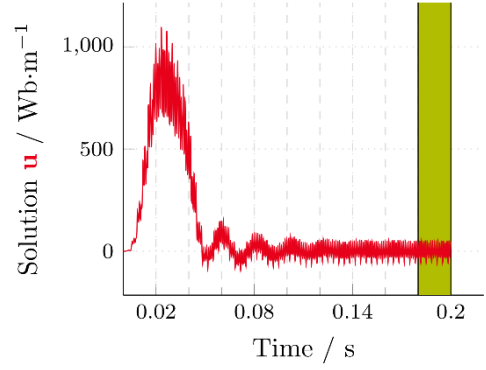

**Fig. 1.** Left: quarter of a four-pole squirrel-cage induction machine model [1, 2]. Right: transient and steady-state evolution of the magnetic vector potential on $[0, 0.2]$ s; the green vertical bar depicts one period $T = 0.02$ s in the steady state.

PP-IC was recently applied to the steady-state analysis of an induction machine in [11], where a significant acceleration was illustrated. Iteration (4) can be written as an operator system over the whole time period $[0, T]$ as

$$\begin{bmatrix} \mathbf{I} & \mathbf{0} & \cdots & -\mathcal{G}_N(\cdot) \\ -\mathcal{G}_1(\cdot) & \mathbf{I} & & \mathbf{0} \\ \vdots & \ddots & \ddots & \vdots \\ \mathbf{0} & \cdots & -\mathcal{G}_{N-1}(\cdot) & \mathbf{I} \end{bmatrix} \begin{bmatrix} \mathbf{U}_0^{k+1} \\ \mathbf{0} \\ \vdots \\ \mathbf{U}_{N-1}^{k+1} \end{bmatrix} = \mathbf{b}^k, \qquad (6)$$

where the right-hand side (RHS) is defined by

$$\mathbf{b}^k = \left[\left[\mathcal{F}_N^k - \mathcal{G}_N^k\right]^\top, \left[\mathcal{F}_1^k - \mathcal{G}_1^k\right]^\top, \ldots, \left[\mathcal{F}_{N-1}^k - \mathcal{G}_{N-1}^k\right]^\top\right]^\top. \qquad (7)$$

We note that system (6) is written in an implicit, possibly nonlinear, form and might have a large size, since the unknown solution vectors at each time instant consist of many degrees of freedom stemming from the spatial discretization. Besides, the system matrix has a block-cyclic structure, typical for periodic problems, which is inconvenient for many linear solvers.

The PP-PC system can be written explicitly for a specific choice of the coarse time-stepping method. In the linear case applying the implicit Euler method

$$\mathbf{C}\left[\mathcal{G}_n\left(\mathbf{U}_{n-1}^{k+1}\right) - \mathbf{U}_{n-1}^{k+1}\right] + \mathbf{K}\mathcal{G}_n\left(\mathbf{U}_{n-1}^{k+1}\right) = \mathbf{f}(T_n) \qquad (8)$$

on the coarse level we obtain an explicit matrix-vector representation of the PP-PC system (6)

$$\underbrace{\begin{bmatrix} \mathbf{Q} & & & -\mathbf{C} \\ -\mathbf{C} & \mathbf{Q} & & \\ & \ddots & \ddots & \\ & & -\mathbf{C} & \mathbf{Q} \end{bmatrix}}_{=:\mathbf{G}} \underbrace{\begin{bmatrix} \mathbf{U}_0^{k+1} \\ \mathbf{U}_1^{k+1} \\ \vdots \\ \mathbf{U}_{N-1}^{k+1} \end{bmatrix}}_{:=\mathbf{U}^{k+1}} = \underbrace{\begin{bmatrix} \mathbf{r}_N^k \\ \mathbf{r}_1^k \\ \vdots \\ \mathbf{r}_{N-1}^k \end{bmatrix}}_{=:\mathbf{r}^k} \qquad (9)$$

with matrices and the RHS given by

$$\begin{aligned} \mathbf{C} &:= \mathbf{M}/\Delta T, \quad \Delta T = T/N, \\ \mathbf{Q} &:= \mathbf{C} + \mathbf{K}, \\ \mathbf{r}_n^k &:= \mathbf{Q}\mathbf{b}_n^k + \mathbf{f}(T_n). \end{aligned} \qquad (10)$$

Note that in the general nonlinear case the matrices on the diagonal of (9) are not the same but depend on the unknown vector. In this case in order to apply the multi-harmonic coarse grid correction an additional treatment of the system matrix is needed, which will be described in our future publications.

In the following section we describe the main ideas of the multi-harmonic approach for efficient solution of the linear PP-PC system (9).

## 3 Multi-harmonic coarse correction for PP-PC

The harmonic balance method (see, e.g., [12, 13, 14, 15]) allows to obtain a periodic solution via calculating its harmonic components. It is based on an expansion into a Fourier series

$$\mathbf{U}_n^{(k+1)} = \sum_{p \in \mathcal{P}} \hat{\mathbf{U}}_p^{(k+1)} \exp\left(\iota \omega_p T_n\right), \; n = 0, \ldots, N-1 \qquad (11)$$

with frequencies of the double-sided spectrum

$$\omega_p = 2\pi p/T, \text{ and } \mathcal{P} := \{-N/2 + 1, \ldots, N/2\}. \qquad (12)$$

Plugging (11) into system (9) and applying the discrete Fourier transform gives an equivalent system in frequency domain:

$$\underbrace{\left(\tilde{\mathbf{F}} \, \mathbf{G} \, \tilde{\mathbf{F}}^H\right)}_{=:\hat{\mathbf{G}}} \hat{\mathbf{U}}^{k+1} = \underbrace{\tilde{\mathbf{F}} \mathbf{r}^k}_{=:\hat{\mathbf{r}}^k}. \qquad (13)$$

The transition from (9) to (13) comes together with an important advantage: it converts the cyclic matrix in (9) to a block-diagonal system matrix of (13). This allows to solve for each harmonic component in parallel. The inverse transform

$$\mathbf{U}^{k+1} = \tilde{\mathbf{F}}^H \hat{\mathbf{U}}^{k+1} \qquad (14)$$

gives the solution in time domain. Note that the discrete Fourier transform and its inverse can be calculated by the fast Fourier transform algorithm, which reduces the computational complexity. Besides, since the structure of the PP-PC system matrix for a particular coarse solver is known, the Fourier transform matrices do not have to be constructed explicitly, but one could solve for each frequency component separately.

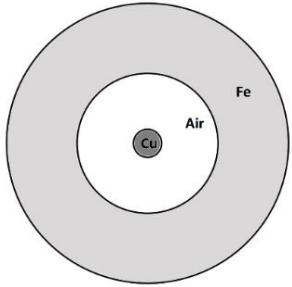

**Fig. 2.** A 2D sketch of the coaxial cable.

## 4 Numerical example

We apply the presented approach to a nonlinear 2D model of a coaxial cable [16]. A sketch of the 2D computational domain is shown in Fig. 2. The eddy current problem (1) is defined on $T = 0.02$ s, with a sinusoidal current input of 50 Hz. The spatial discretization is performed with the finite element method, using 2269 degrees of freedom.

We split the time interval into $N = 20$ subintervals. The fine propagator uses time step size $\delta T = 10^{-5}$ s, and the coarse step size is $\Delta T = 10^{-3}$ s. The relaxed PP-IC iteration [8] required 19 iterations to reach tolerance $< 10^{-6}$. The fixed point iteration from [8] required 5 iterations, whereas PP-PC with the multi-harmonic coarse grid correction introduced in this paper converged in 3 iterations. Besides, the solution of the PP-PC system is accelerated via parallelization of calculations in frequency domain, i.e., the effective effort is solving one linear system of a convenient structure instead of solving the 20 times larger cyclic system.

## 5 Conclusion

This paper presents a parallel-in-time approach for the efficient treatment of time-periodic problems. A multi-harmonic correction on the coarse grid reduces the complexity of the solutions of the resulting linear system by transforming a cyclic system matrix into a block-diagonal form. This brings an additional acceleration using parallelization on the coarse grid. The method was tested with a 2D nonlinear coaxial cable model. It was seen that the introduced algorithm outperformed the existing approaches due to its reduced complexity and a smaller computational cost. Future research will focus on more realistic applications and describe the nonlinear case in detail.

## Acknowledgements

This work is supported by the Excellence Initiative of the German Federal and State Governments, the Graduate School of Computational Engineering at TU Darmstadt, the BMBF (05M2018RDA) and DFG (SCHO1562/1-2).